\documentclass[10pt,showpacs,preprintnumbers,amsmath,amssymb]{revtex4}

\usepackage{epsf}
\usepackage{graphicx}  
\usepackage{dcolumn}   
\usepackage{bm}        

\newcommand{\be}{\begin{equation}}
\newcommand{\en}{\end{equation}}
\newcommand{\ba}{\begin{eqnarray}}
\newcommand{\ea}{\end{eqnarray}}

\begin{document}

\title{New mechanism to cross the phantom divide}

\author{Yunshuang Du}
\author{Hongsheng Zhang}
\author{Xin-Zhou Li  }
\affiliation{Shanghai United Center for Astrophysics (SUCA),
 Shanghai Normal University, 100 Guilin Road, Shanghai 200234,
 P.R.China}
 \date{ \today}

\begin{abstract}
Recently,  type Ia supernovae data appear to support a dark energy
whose equation of state $w$ crosses $-1$, which is a much more
amazing problem than the acceleration of the universe. We show that
it is possible for the equation of state to cross the phantom divide
by a  scalar field in the gravity with an additional inverse
power-law term of Ricci scalar in the Lagrangian. The necessary and
sufficient condition for a universe in which
  the dark energy can cross the phantom divide is obtained. Some analytical solutions with $w<-1$ or
  $w>-1$ are obtained. A minimal coupled scalar with different potentials, including quadratic,
  cubic, quantic, exponential and logarithmic potentials are investigated via numerical methods,
   respectively. All these potentials lead to the crossing behavior. We show that it is a
   robust result which is hardly dependent on the concrete form of the potential of the scalar.

\end{abstract}

\pacs{95.36.+x 98.80.Qc 04.20.Cv} \keywords{Cosmic acceleration, scalar field, modified gravity}

\maketitle
\section{Introduction}
Over the course of the past decade, the evidence for the acceleration
of the the universe \cite{acce} is one of the most striking cosmological
 discoveries. A  number of approaches, namely the existence of dark energy,
  have been adopted to try and explain the remarkable observation. An
  incomplete list includes: quintessence, phantom, k-essence, modified gravity, and
  so on \cite{review}. The quintessence field models correspond to
   an equation of state $w\geq-1$. The region where the equation of
   state is less than $-1$ is typically refer to as a presence due to something called  phantom dark energy.
   Furthermore, in the wake of the more accurate data, a surprising result emerges: the resent study of the
   type Ia supernovas data show that the time varying dark energy provides a better fit than the cosmological
    constant, in addition, especially, the equation of state parameter $w$, defined as the ratio of pressure to
    energy density, may cross the phantom divide $w=-1$. This crossing to the phantom region is neither possible
     for an ordinary minimally coupled scalar field nor for a phantom field. It clearly shows that this transition
      is difficult to be realized by only a single-field models of dark energy. In this article, we try to explore
      new ways and means to obtain the consistency for the theoretical outcome with the present observation.

The contribution of the matter content of the universe is represented by the energy momentum tensor on the right
 hand side of Einstein equations, whereas the left hand side is represented by pure geometry.
 Thus, there are two ways to give rise to current acceleration: (i) supplementing the energy momentum
  tensor by an exotic  matter with negative pressure such as a cosmological constant or a scalar field;
   (ii) modifying the geometry itself. Since the dark energy models mentioned previously have not gotten fully satisfactory
     explanations of cosmic acceleration, it  is worthwhile considering the possibility that acceleration
      of universe is not due to some kind of stuff, but rather is caused by modified gravitation. Lately,
       there is a very interesting operation on the general relativity: a modification to the Einstein-Hilbert
       action by involving new terms of inverse powers of the curvature scalar (in the following text we call
        it inverse-R gravity), of the form $\sqrt{-g}R^{-n}$(with $n>0$) \cite{troden}.
        Terms of this form would become important in the late universe and can bring about self-accelerating
          solutions, supplying a purely geometric candidate for dark energy. But, as shown in \cite{troden},
          the effect of extra term $\sqrt{-g}R^{-n}$ equals to a scalar field with  a smooth positive potential.
          Therefore, the EOS of dark energy still fails to cross the phantom divide yet.   Given the challenge of this
           problem, we  investigate a novel model to pursue the
           challenge.    In general $f(R)$ gravity (even without scalar field),    the phantom divide crossing behavior of dark energy is shown in
           \cite{odin1,odin2},
         and  $R^2$-term can yield a transient crossing \cite{odin1,odin2}.

In the next section we shall make a brief review of inverse-R
gravity and its relation to the scalar theory in Einstein gravity.
In the section III we investigate the mechanism for a single scalar
to cross the phantom divide by an analytical method. In section IV,
we present the numerical results of the crossing behavior of a
scalar with different potentials in frame of inverse-R gravity in
detail. And in Section V, we will present the main conclusion and
some discussions.

\maketitle
\section{inverse-R gravity}

 There are varies modifications of Einstein-Hilbert action. Basically, we classify them into two categories:
 one is ultraviolet modification which will be important in the high curvature region (early universe)
  while the other is infrared modification which plays significant role in the low curvature region (late universe).
  The ultraviolet modification has been investigated widely in the inflation models, for example the starobinsky inflation \cite{star}.
   Recently, the infrared modification arouses much interests because of the discovery of cosmic acceleration.
    The inverse-R gravity is one of the leading model in the model with infrared
    modifications. It is noted that the inverse-R gravity is equivalent to a class of
 scalar tensor theory with $\omega=0$, which is not compatible with solar system observations \cite{chiba}. However, the physical meaning
 of this approach is not clear enough as the first sight. For example, it is shown that a specific $f(R)$ gravity implying cosmic
 acceleration does not physically equal to scalar tensor theory, though it do equal scalar tensor theory mathematically \cite{noneq}.
 Thus, it is still sensible to study  modified gravity and scalar tensor theory, respectively.

 Moreover,
 a scalar
 curvature squared term to the action can save
  the inverse-R gravity and help it to pass the observations in solar system
  \cite{noji}.
     We study the behavior of a scalar field in the inverse-R gravity with $R^2$ correction, whose action $S$ reads,
 \be
\label{action} S =\frac{m_{pl}^2}{2}\int d^4 x\,
\sqrt{-g}\left(f(R)\right) +\int d^4 x\, \sqrt{-g}\, ({\cal L}_M
+{\cal L}_{phi}). \en
  Here $m_{pl}$ is Planck mass, $g$ denotes the determinant of the metric, $R$ marks the Ricci scalar,
   $\alpha, \beta$ are constants with mass dimension, ${\cal L}_{ M}$ is the matter Lagrangian,

  \be
 f(R)=R-\frac{\alpha^4}{R}+\beta^2R^2,
 \label{fr}
 \en
   and
 ${\cal L}_{\phi}$ labels the Lagrangian of a scalar field with potential $V(\phi)$,
 \be
 {\cal L}_{\phi}=-\frac{1}{2}\partial_\mu \phi \partial^\mu \phi-V(\phi).
 \en
 In this set-up, $\beta$ is at the scale of inflation and $\alpha$ is at the scale of the present universe, hence $\beta\gg\alpha$.
  Note that $R^2$ term aids the original inverse-R gravity to pass
 the tests in solar system and to ensure the universe undergoes a
 matter-dominated era, though we should ensure its effects in the late universe are tiny.
 The action (\ref{action}) yields field equation,
 \be
  \label{fd}
 -g_{\mu\nu}f+2R_{\mu\nu}\frac{\partial f}{\partial
 R}+2g_{\mu\nu}\square \frac{\partial f}{\partial
 R} -2\triangledown_{\mu}\triangledown_{\nu}\frac{\partial f}{\partial
 R}
  =\frac{T_{\mu\nu}^M+T_{\mu\nu}^{\phi}}{m_{pl}^2},
\en
 where $T_{\mu\nu}^M, T_{\mu\nu}^{\phi}$ denote the energy-momentums of matter and scalar field, respectively. In the present article,
  we suppose that matter is just dust. By assuming an  FRW universe with scale factor $a$,
  \be
  ds^2=-dt^2+a^2d\Sigma^2,
  \en
 where $d\Sigma^2$ denotes the metric of a 3-dimensional maximally symmetric space,
  we obtain the friedmann equation corresponding to (\ref{fd}),
  \be
  \label{fried}
  H^2+\frac{k}{a^2} -\frac{F(H,\dot{H},\ddot{H})}{3m_{pl}^2} = \frac{\rho_M+\dot{\phi}^2/2+V}{3m_{pl}^2}\ ,
 \en
    where $H\triangleq {\dot{a}/ a}$ is the Hubble parameter, a dot denotes derivative with respect to the cosmic time $t$, and
     \be
 F=\frac{m_{pl}^2\alpha^4}{12({\dot
H}+2H^2)^3}\left(2H{\ddot H}  + 15H^2{\dot H}+2{\dot
H}^2+6H^4\right)-18(6H^2\dot{H}-\dot{H}^2+2H\ddot{H})
\frac{m_{pl}^2}{\beta^2}.
  \label{F}
  \en

    Another equation to close this system is the
    acceleration equation, which can be replaced by the equation of motion for $\phi$,
    \be
    \ddot{\phi}+3H\dot{\phi}+\frac{dV}{d\phi}=0.
    \label{eom}
    \en

   In this construction the scalar field together with the extra geometric term plays the role of dark energy.   First we present a concise
   note on the definition of dark energy.  In the inverse-R gravity theory,
   there is a surplus $F$-term in the modified Friedmann equation (\ref{fried}).
    However, almost all observed properties of dark energy are
    obtained in frame of general relativity.
    To explain the the observed evolving  EOS of the effective dark
    energy,  we introduce the concept ``equivalent
 dark energy" or ``virtual dark energy" in the modified gravity
 models \cite{reviewcross}.  We derive the density of virtual dark energy caused by the non-minimal coupled scalar by comparing the modified Friedmann equation in
  the inverse-R gravity with the standard Friedmann equation in general
  relativity.
   The generic Friedmann equation in the
 4-dimensional general relativity can be written as
 \be
 H^2+\frac{k}{a^2}=\frac{1}{3m_{pl}^2} (\rho_{dm}+\rho_{de}),
 \label{genericF}
 \en
 where the first term of RHS in the above equation represents the dust matter and the second
 term stands for the dark energy. Comparing (\ref{genericF})
 with (\ref{fried}), one obtains the density of virtual dark
 energy in STG,
 \be
 \rho_{de}=\rho_{\phi}+F,
 \label{rhode}
 \en
 where
 \be
 \rho_{\phi}=\frac{ \dot{\phi}^2}{2}+V.
 \en

 We see that the $F$-term makes a key difference from the standard model. In the next section, we shall study the dynamics
 of a universe in inverse-R gravity, especially the dynamics of the virtual dark energy (dark energy for short).

 \section{exact solutions in inverse-R cosmology}
 To find exact solutions is an important but difficult topic in such a high non-linear system (\ref{fried}) and (\ref{eom}).
 In this section we study some exact solutions to explore the mechanism to cross the phantom divide for the dark energy in inverse-R
 cosmology. In this section we only consider an ideal case, that is,
 $\beta=0$, to see some quanlative behavior of the late universe in inverse-R gravity. And we leave the more realistic case with $R^2$-term in
 the numerical analysis.
 \subsection{Power-law solution}
 In a spatially flat universe, the Friedmann equation (\ref{fried}) reduces to
  \be
  \label{fried1}
  H^2+\frac{k}{a^2} -\frac{F(H,\dot{H},\ddot{H})}{3m_{pl}^2} = \frac{\dot{\phi}^2/2+V}{3m_{pl}^2}\ ,
 \en
 during the dark energy dominated stage. Square-law solution arises when the potential is chosen to take the exponential form,
 \be
 V=m_e^4e^{\frac{\phi}{m_{pl}}},
 \en
 where $m_e$ is a constant. With this potential, (\ref{fried1}) and (\ref{eom}) have a particular solution as follows,
 \be
 a=a_0t^2,
 \en
 \be
 \phi=2m_{pl}\ln \left(\frac{m_e^2}{10}\frac{t}{m_{pl}}\right).
 \en
It is easy to calculate that $w=-2/3$ in this case.
 One sees that this universe is ``uniformally accelerating", since $\ddot{a}=$constant. An interesting point in this
 uniformally accelerating universe is that the extra term $F$ vanishes just in time, and hence it is the solution with which we are familiar in standard general relativity by coincidence.
 \subsection{De Sitter solution}
 The dynamical system of the scalar filed with canonical and
 non-canonical Lagrangian has been widely studied \cite{hl1}, among
 which the global structure of the phase space has been investigated
 and various critical points and their physical significance have been identified and manifested. We
 have shown that if the potential of the model with kinetic energy
 (quintessence) has a non-vanishing minimum \cite{hl2}, or the
 potential of a model with negative energy (phantom) has
 non-vanishing maximum \cite{hl3}, the dynamical system of the model
 admits late-time attractor solutions corresponding to $w_{\phi}=-1$
 and $\Omega_{\phi}=1$, where $\Omega$ denotes the fraction of a
 stuff of the universe.

 For a class of model that admits late-time de Sitter attractor, we have a solution,
 \be
 a=a_0e^{\left[V_0/6m_{pl}^2+(V_0^2/m_{pl}^4+\alpha^4/12)^{1/2}/2\right]^{1/2}(t-t_0)},
\label{attra}
 \en
 \be
 \phi=\phi_0.
 \en
  From (\ref{attra}) and (\ref{F}), we derive
  \be
  F=\frac{\alpha^4m_{pl}^4}{8\left[V_0/3m_{pl}^2+(V_0^2/9m_{pl}^4+\alpha^4/12)^{1/2}\right]}.
  \en
  Thus $\frac{d\rho_{\phi}}{da}=\frac{dF}{da}=0$ and the crossing
  phenomenon cannot happen near the late-time de Sitter attractor.
  \subsection{Big rip solution}
 Present observation data indicate the possibility of dark energy
 with $w<-1$, dubbed as phantom \cite{hl4}. In the ordinary case,
 the realization of $w<-1$ could not be achieved. Unfortunately, it
 suffers from the well known quantum stability problem. Though there
 are a few discussions to evade the stability problem for phantom
 model, we are far from ``solving it" \cite{sta}. A good $w<-1$ model should
 avoid the negative kinetic energy as much as possible.

 Under the potential,
 \be
 V(\phi)=V_0\left(-\int_{\frac{\phi_0}{m_{pl}}}^{\frac{\phi}{m_{pl}}}\frac{du}{1+\gamma[{
 \cal I}(u)]^4}\right),
 \en
 where ${\cal I}(u)$ is the inverse function of the function
 $u=m_{pl}(\beta \ln v+\frac{3\gamma v^4}{2})$, and
 \be
 \beta=7+\sqrt{73},
 \en
 \be
 \gamma=\frac{\alpha^4}{200(25+\sqrt{73})},
 \en
 we have a big rip solution,
 \be
 a=a_0(t-t_{br})^{-2},
 \en
 \be
 \phi=m_{pl}\left[\beta\ln (t_{br}-t)+\frac{3\gamma
 (t-t_{br})^4}{2}\right],
 \en
 where $t_{br}$ is the rip epoch. For this
 solution, we have
 \be
 w=-\frac{4}{3}.
 \en
 \subsection{The crossing mechanism}
  Since there are both $w>-1$ and $w<-1$ solutions, the crossing
  phenomenon is bounded to happen in our model.

   Before studying the crossing behavior in this model, we say a bit more on the relation between density evolution and the equation of state.
   In fact, we can know that dark energy behaves as quintessence, vacuum or phantom only from its  rate of change of the density. We reach this point from the following arguments.
 Since the dust matter obeys the continuity equation
 and the Bianchi identity keeps valid, dark energy itself satisfies
  the continuity equation
 \be
 \frac{d\rho_{de}}{dt}+3H(\rho_{de}+p_{eff})=0,
 \label{contieff}
 \en
 where $p_{eff}$ denotes the effective pressure of the dark energy.
 And then we can express the equation of state for the dark
 energy as
   \be
  w=\frac{p_{eff}}{\rho_{de}}=-1-\frac{a}{3\rho_{de}}\frac{d \rho_{de}}{d
  a}.
  \label{wde}
   \en
   From the above equation we find that the behavior of $w_{de}$ is determined by the term $\frac{d \rho_{de}}{d
  a}$. $\frac{d \rho_{de}}{d
  a}=0$ (cosmological constant) bounds phantom and quintessence. More
  intuitively, if $\rho_{de}$ increases with the expansion of the universe, the dark energy
  behave as phantom;  if $\rho_{de}$ decreases with the expansion of the universe, the dark energy
  behave as quintessence; if $\rho_{de}$ decreases and then increases, or increases and then
  decreases, we are certain that EOS of dark energy crosses phantom
  divide. A more important reason why we use the density to describe
  property of dark energy is that the density is
  more closely related to observables, hence is more tightly
  constrained for the same number of redshift bins used \cite{wangyun}.
 With data accumulation, observations which favor dynamical dark
   energy become more and accurate.   Furthermore, a significant possibility
   appears recently: the EOS of dark energy may cross $-1$ (phantom divide) \cite{vari}, which is a serious challenge for theoretical
   physics. The theoretical explore of the crossing phenomenon was proposed in \cite{cros}.
   This interesting topic is under intensively studying very recently \cite{cross}, see also \cite{reviewcross} and for some earlier references therein.

  (\ref{wde}) can be rewritten as
  \be
  w=-1-\frac{1}{3H\rho_{de}}(\dot{\rho}_{\phi}+\dot{F}).
  \en
  The behaviors of $F$ and $\rho_{\phi}$, increase or decrease with
  expansion of the universe, depend on the rate of expansion. A
  simple calculation presents,
  \be
  \dot{F}=\frac{Hm_{pl}^2\alpha^4}{6(2H^2+\dot{H})^4}
  \left[\ddot{H}(2H^2+\dot{H})+3\ddot{H}(2H^2-7H\dot{H}-\ddot{H})+
  3\dot{H}(\dot{H}^2-16H^2\dot{H}-4H^4)\right].
  \label{fdot}
  \en
  For a power-law universe,
  \be
  a=a_0\left(\frac{t}{t_0}\right)^\xi,
  \label{factor}
  \en
  where $t_0$ is the present age of the universe, $\xi$ is a
  constant. $\xi>0$ denotes an expanding universe; $\xi>1$ implies
  an accelerating one; $\xi>2$ implies $\dddot{a}>0$, hence $F>0$
  and $\dot{F}>0$, which presents a possibility to cross the phantom
  divide for the dark energy. Substituting (\ref{factor}) into
  (\ref{fdot}), we derive,
  \be
  \dot{F}=\frac{(\xi-2)m_{pl}^2\alpha^4}{2\xi^3(2\xi-1)^2}.
  \en
  In the $\xi=\frac{11+\sqrt{73}}{2}$ case, we have
  $\dot{F}_{max}=1.75\times 10^{-6}$. If $\xi>>1$, we have
  $\dot{F}<<1$. It is consistent that $\dot{F}=0$ in the de Sitter
  expansion, because de Sitter expansion is faster than any
  power-law accelerating expansion.

  From (\ref{eom}) and (\ref{fried1}), we find that
  \be
  \dot{H}=\frac{-3H\dot{\phi}^2+\dot{F}}{6m_{pl}^2H},
  \en
  \be
  \ddot{H}=\frac{-3\dot{H}\dot{\phi}^2-6H\dot{\phi}\ddot{\phi}+\ddot{F}-6m_{pl}^2\dot{H}^2}{6m_{pl}^2H}.
  \en

  The universal argument is that the crossing phenomenon happens at
  $t=t_c$ iff $\dot{H}_{t=t_c}=0$ and $\ddot{H}_{t=t_c}\neq 0$. In
  the quintessence case in standard model, ie, $F\equiv 0$, $\dot{H}_{t=t_c}=0$ implies $\ddot{H}_{t=t_c}\neq
  0$. Therefore, the crossing behavior never happens. In inverse-R
  gravity, $\dot{H}_{t=t_c}=0$ and $\ddot{H}_{t=t_c}\neq 0$ iff,
  \be
  \dot{F}=3H\dot{\phi}^2,
  \en
  and
  \be
  \ddot{F} \neq 3H\dot{\phi}^2.
  \en
  We now have proved two theorems as follows,

  $Theorem~ 1$ The crossing phenomenon happens iff
  $\dot{F}=3H\dot{\phi}^2$, $\ddot{F} \neq
  6H\dot{\phi}\ddot{\phi}$, $\dot{\phi}\neq 0$ and $\ddot{\phi}\neq
  0$.

  $Theorem ~2$ Iff $\dot{F}<3H\dot{\phi}^2$, we have the model $w<-1$;
  iff $\dot{F}>3H\dot{\phi}^2$, we have the model with $w>-1$.

 \section{Numerical examples}

  In the above section, we discuss the evolution of the universe in the
  scalar dominated stage via a analytical way. For a realistic universe, the pressure-less dust is a
  necessary component. However, it difficult to study the evolution of the universe when we introduce a dust in inverse-R gravity with $R^2$ corrections.
    In this section, we investigate the dynamics of scalars with five different potentials, including quadratic, cubic,
    quantic, inverse power-law, exponential and logarithmic potentials in a universe with dust component via numerical methods, respectively.
    In the following text, we always assume a spatially flat universe.

   \subsection{quadratic potential}
   Quadratic potential is the most widely-investigated potential in field theory, which represents a mass term. Any potential around its minimum (if it has a minimum) can be treated as quadratic potential. In this subsection we study the case of quadratic potential,
   \be
   V=m^{2}_2\phi^{2},
   \en
   where $m$ is a parameter with dimension of mass.

    For convenience, we introduce two dimensionless variables
\be
  x\triangleq H/H_{0},    y\triangleq \phi/M_{p},
 \en
                     with which we rewrite Friedman equation (\ref{fried}) and the equation of motion (\ref{eom}) for $\phi$ into the following form,
\be
     x^{2}
     =\Omega_{m0}e^{-3s}+\frac{1}{6}x^{2}y'^{2}+\frac{1}{3}k_2y^{2}+\frac{b(4x^{2}x'^{2}+2x^{3}x''+15x^{3}x'+6x^{4})}{36(xx'+2x^{2})^{3}}
     +18c(6x^3x'+x^2x'^2+2x^3x''),
     \en

\be
     0= xx'y'+x^{2}y''+3x^{2}y'+2k_2y,
      \en
     where a prime denotes the derivative with respect to the so-called e-folds $s\triangleq \ln a$,
     $b\triangleq \frac{\alpha^{4}}{H_{0}^{4}},~c\triangleq \frac{H_0^{2}}{\beta^{2}}$ is a dimensionless constant, the coefficient  $k_2\triangleq \frac{m^{2}_2}{H_{0}^{2}}$ and

 \be
  \Omega_{m0}\triangleq \frac{\rho_{M0}}{H_{0}^2m_{pl}^2}.
  \en
  Here  $H_{0}$ is the current Hubble parameter, $\rho_{M0}$ represents the current density of
  dust. Generally, $c$ is a tiny number since the inflation scale
  $\beta$ is much higher than $H_0$. Hence, the effects of $R^2$-term in (\ref{fr})
  can be omitted in late universe, although it becomes important in
  the early universe and structure formation era. This also implies
  that the exact solutions discussed in the last section make
  physical sense in the late universe.

    Though we get the sufficient condition for the $F$-term to help the dark energy to cross the phantom divide, it is difficult to apply this condition before we obtain the exact solution. However in a universe with multiple components, one hardly obtain an exact solution.     Under this situation, we present a numerical result about the dark energy density in Fig.1. For the sake of convenience, we introduce the dimensionless density  as below,
    \be
     \beta= \frac{\rho_{de}}{3H_{0}^2m_{pl}^2}
     =\frac{1}{6}x^{2}y'^{2}+\frac{1}{3}k_2y^{2}+\frac{b(4x^{2}x'^{2}+2x^{3}x''+15x^{3}x'+6x^{4})}{36(xx'+2x^{2})^{3}}+18c(6x^3x'+x^2x'^2+2x^3x'').
     \en

\begin{figure}
\centering
\includegraphics[scale=1.]{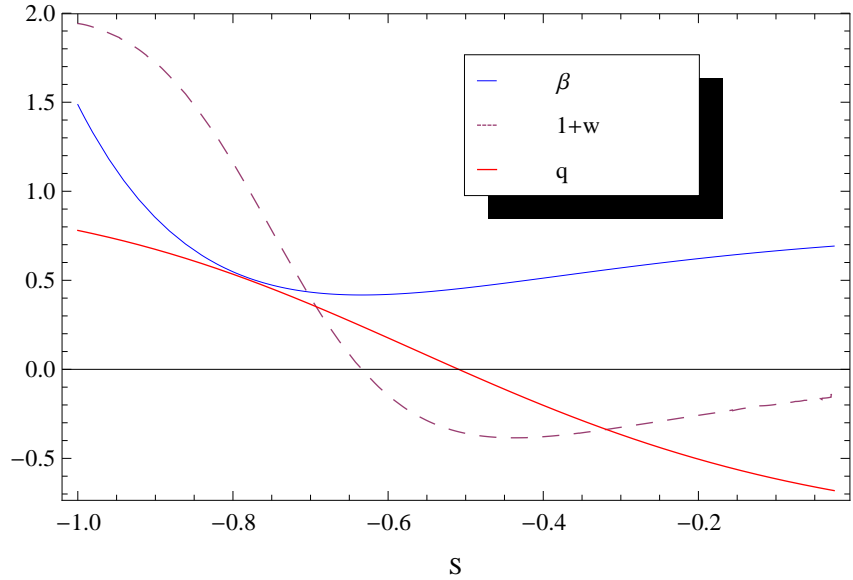}
\caption{ $\beta$, $w$ and $q$ as functions of $s$. This figure is
for  $V=m^{2}\phi^{2}$, in which we set $\Omega_{m0}=0.3$, $k_2 = 1,
b = 20, c=10^{-116}$, and initial values for arguments: $x_{0}=1,
y_{0} = 0, x_{0}' = -0.3, y_{0}' = 0.$} \label{rho2}
\end{figure}

    The most important parameter from the aspect of observation is the deceleration parameter $q$, which carries the total effects of the cosmic fluids,
\be
    q=-\frac{\ddot{a}a }{\dot{a}^{2}}=-1-\frac{x'}{x}.
\en
 We also displays $q$ in Fig. (\ref{rho2}).

   From Fig. (\ref{rho2}), obviously, the equation of state of dark energy crosses -1. Simultaneously, the deceleration parameter is consistent with observation. As is known to all, the equation of state of a single scalar in standard general relativity never crosses the phantom divide, the $F$-term, induced from the  $R^{-1}$ term in the Lagrangian, plays an essential role in this crossing.

 \subsection{cubic potential}
  Next, we study the potential $V=m_3\phi^{3}$. This potential has been investigated in classical theory, especially in the dynamics of an oscillator.
  As for $V=m_3\phi^{3}$, the Friedman equation (\ref{fried}) and the equation of motion (\ref{eom}) become

\be
  x^{2}=\Omega_{m0}+\frac{1}{6}x^{2}y'^{2}+\frac{1}{3}k_3y^{3}+\frac{b(4x^{2}x'^{2}+2x^{3}x''+15x^{3}x'+6x^{4})}{36(xx'+2x^{2})^{3}}
  +18c(6x^3x'+x^2x'^2+2x^3x''),
\en

\be
   0= x
   x'y'+x^{2}y''+3x^{2}y'+3k_3y^{2},
  \en
 where $k_3=m_{pl}m_3/H^2_0$.

   The results in  this situation is presented in Fig.2. One can see easily the crossing behavior as mentioned above.

\begin{figure}
\centering
\includegraphics[scale=1.]{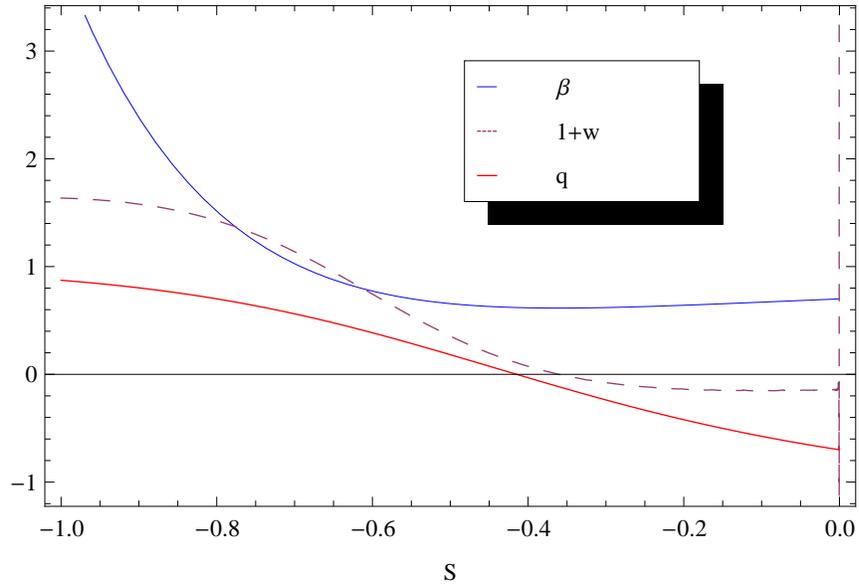}
\caption{This figure is for the evolutions of $\beta$, $w$, and $q$
for  $V=m_3\phi^{3}$, in which $\Omega_{m0}=0.3$, and initial values
for coefficients: $k_3 = 6, b = 20$, $x_{0}=1, y_{0} = 0, x_{0}' =
-0.3, y_{0}' = 0, c=10^{-116}$ }
\end{figure}

 \subsection{quantic potential}
 In this subsection we explore the dynamics of the universe for a scalar with $V=c\phi^{4}$(c is a dimensionless constant) in inverse-R gravity. The quantic potential is
 extremely important in modern physics. Due to Higgs mechanism, every massive particle get mass through a quantic potential.
   Then, for  $V=c\phi^{4}$, the two principle equations become
 \be
 x^{2}=\Omega_{m0}+\frac{1}{6}x^{2}y'^{2}+\frac{1}{3}k_4y^{4}+\frac{b(4x^{2}x'^{2}+2x^{3}x''+15x^{3}x'+6x^{4})}{36(xx'+2x^{2})^{3}}+18c(6x^3x'+x^2x'^2+2x^3x''),
  \en

 \be
  0=xx'y'+x^{2}y''+3x^{2}y'+4k_4y^{3},
\en
  in which $k_4=cm_{pl}^2/H_0^2$.

      We give the results in Fig.3.

  \begin{figure}
\centering
\includegraphics[scale=1.]{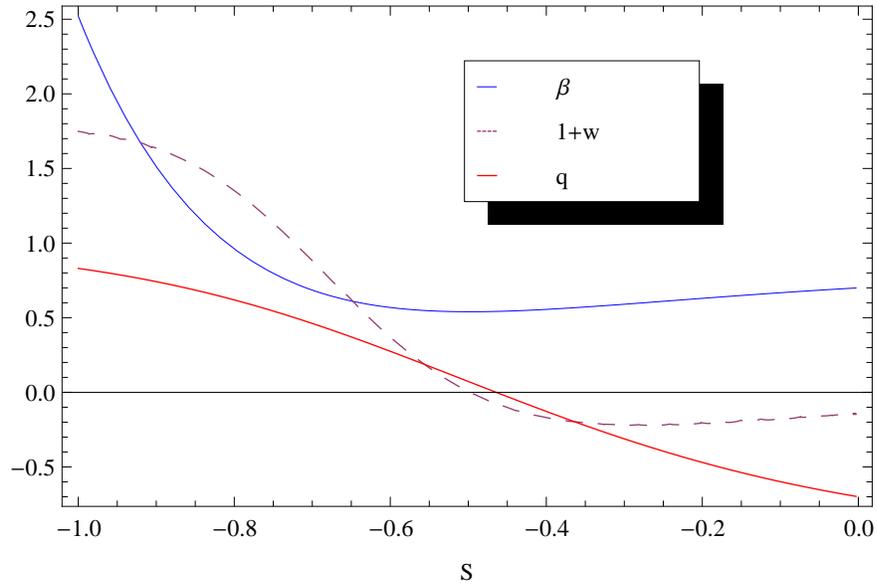}
\caption{ $\beta$, $w$ and $q$ as functions of s. In this figure we
set  $V=c\phi^{4}$, $\Omega_{m0}=0.3$, and initial values for
coefficients: $k_4 = 11, b = 20$, $x_{0}=1, y_{0} = 0, x_{0}' =
-0.3, y_{0}' = 0, c=10^{-116}$}
\end{figure}

\subsection{exponential potential}
      In this subsection, we study the exponential potential. In the standard inflation models, the exponential potential is an important
example which can be solved exactly in the standard model. In
addition, we know that such exponential potentials of scalar
fields occur naturally in some fundamental theories such as
string/M theories.  We present  the results  about $V=m_e^4e^{\frac{\phi}{m_{pl}}} $  in following content.

      Concerning $V=m_e^4e^{\frac{\phi}{m_{pl}}} $, the equations (\ref{fried}) and (\ref{eom}) to be

 \be
      x^{2}=\Omega_{m0}+\frac{1}{6}x^{2}y'^{2}+\frac{1}{3}k_ee^{y}+\frac{b(4x^{2}x'^{2}+2x^{3}x''+15x^{3}x'+6x^{4})}{36(xx'+2x^{2})^{3}}+18c(6x^3x'+x^2x'^2+2x^3x''),
             \en

 \be
                   0=xx'y'+x^{2}y''+3x^{2}y'+k_ee^{y},
\en
 where $k_e=m_e^4H_0^{-2}m_{pl}^{-2}$.

 The results are given by Fig.4.

 \begin{figure}
\centering
\includegraphics[scale=1.]{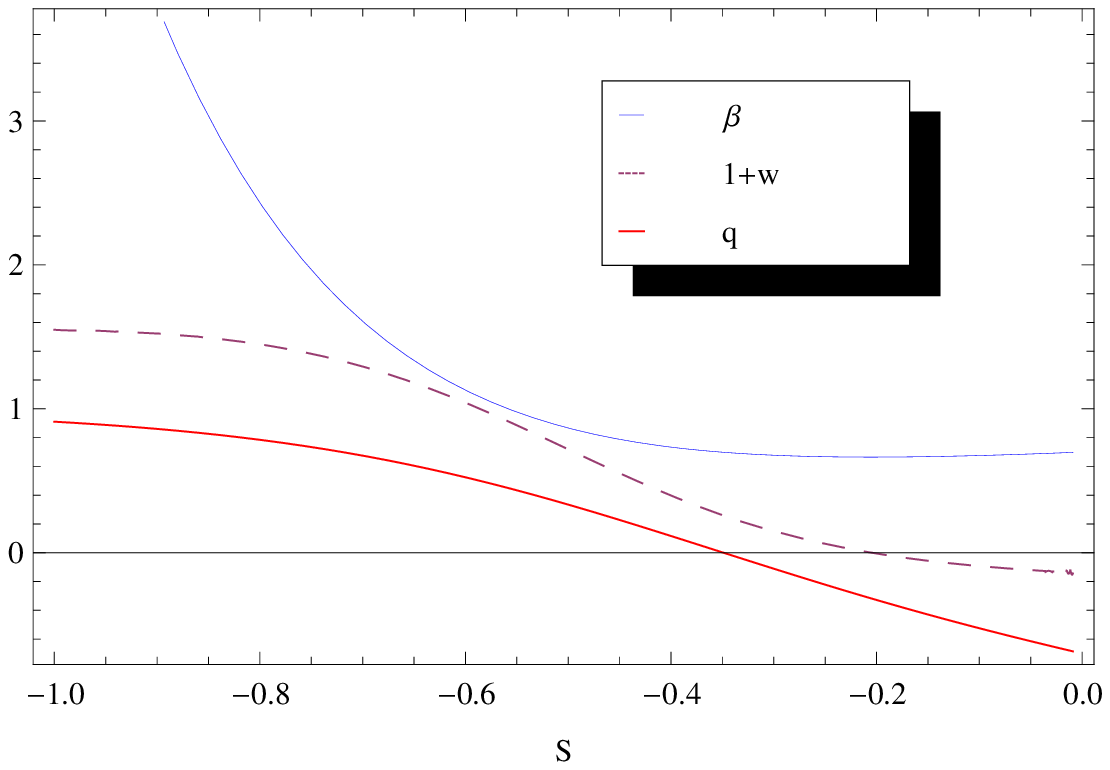}
\caption{This figure is for  $V=m_e^4e^{\frac{\phi}{m_{pl}}} $,
$\Omega_{m0}=0.3$, and initial values for coefficients: $k_e = 1, b
= 10$, $x_{0}=1, y_{0} = 0, x_{0}' = -0.3, y_{0}' = 0, c=10^{-116}.$
}
\end{figure}

  \subsection{logarithmic potential}

 The logarithmic potential is not widely studied in particle physics. To fit parts of elementary particle mass
 spectra by involving logarithmic potentials comes into physicists' view only in recent decades \cite{lopo}.
  In this paper, we study it phenomenologically. With regard to  $V=m_l^4\ln\frac{\phi}{m_{pl}}$, the corresponding equations (\ref{fried}) and (\ref{eom}) will be

  \be
    x^{2}=\Omega_{m0}+\frac{1}{6}x^{2}y'^{2}+\frac{1}{3}k_l\ln{y}+\frac{b(4x^{2}x'^{2}+2x^{3}x''+15x^{3}x'+6x^{4})}{36(xx'+2x^{2})^{3}}+18c(6x^3x'+x^2x'^2+2x^3x''),
      \en

  \be
         0=xx'y'+x^{2}y''+3x^{2}y'+\frac{k_l}{y},
  \en
  where $k_l=m_l^4H_0^{-2}m_{pl}^{-2}$.

We present the results in Fig.5.

\begin{figure}
\centering
\includegraphics[scale=1.]{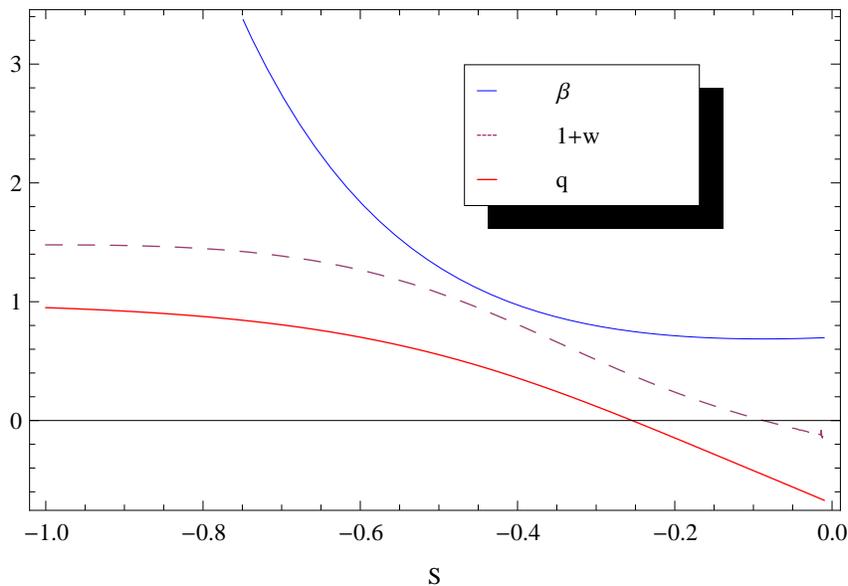}
\caption{This figure for  $V=m_l^4\ln\frac{\phi}{m_{pl}}$, $\Omega_{m0}=0.3$, and initial values for coefficients:
 $k_l = 1, b = 20$, $x_{0}=1, y_{0} = 1, x_{0}' = -0.3, y_{0}' = 0, c=10^{-116}.$}
\end{figure}

      From figs. 1-5, one sees that all the $w$ of the virtual dark energy with different potentials are successfully to make the equation of state of dark energy cross $-1$. And the 3 curves in all the figures have a similar shape, which implies that the crossing behavior is determined by the extra geometric term. Consequently, we have to say that it is geometric property of the inverse-R gravity itself, which is independent of the individual potentials.

     \maketitle
\section{conclusion}

In conclusion, the results of this paper  demonstrate convincingly
that it is possible to realize the crossing $w=-1$ for the equation
of state of the dark energy by a single scalar. We find the
necessary and sufficient condition for a universe in which the dark
energy cross the phantom divide in inverse-R gravity.   And then we
investigated different potentials to minimally coupled scalar field
$\phi$ , including quadratic, cubic, quantic, exponential,
logarithmic potentials in inverse-R gravity with $R^2$ correction.
And the results state clearly that different potentials lead to the
crossing behavior, respectively. Therefore, we  conclude that it is
a robust property of inverse-R gravity with $R^2$ correction, not
controlled by a special potential.

{\bf Acknowledgments.}
 We thank the anonymous referee for his several valuable suggestions. This work is supported by National Education Foundation of China under grant No. 200931271104,
 Shanghai Municipal Pujiang grant No. 10PJ1408100, and National Natural Science Foundation of China under Grant No. 11075106.

\end{document}